\documentclass[showpacs,preprintnumbers,amsmath,amssymb,11pt]{revtex4}
\usepackage{graphicx}

\begin{document}

\title{Constraint-preserving boundary conditions in the 3+1 first-order approach}

\author{C.~Bona and C.~Bona-Casas}
\affiliation{Departament de Fisica, Universitat de les Illes
Balears, Palma de Mallorca, Spain.\\
Institute for Applied Computing with Community Code
(IAC$^{\,3}$)}.

\pacs{04.25.Dm, 04.20.Cv}

\begin{abstract}
A set of energy-momentum constraint-preserving boundary conditions
is proposed for the first-order Z4 case. The stability of a simple
numerical implementation is tested in the linear regime (robust
stability test), both with the standard corner and vertex
treatment and with a modified finite-differences stencil for
boundary points which avoids corners and vertices even in
cartesian-like grids. Moreover, the proposed boundary conditions
are tested in a strong field scenario, the Gowdy waves metric,
showing the expected rate of convergence. The accumulated amount
of energy-momentum constraint violations is similar or even
smaller than the one generated by either periodic or reflection
conditions, which are exact in the Gowdy waves case. As a side
theoretical result, a new symmetrizer is explicitly given, which
extends the parametric domain of symmetric hyperbolicity for the
Z4 formalism. The application of these results to first-order
BSSN-like formalisms is also considered.
\end{abstract}

\maketitle

\section{Introduction}

Constraint-preserving boundary conditions is an active research
topic in Numerical Relativity. During the first half of this
decade, many conditions have been proposed, adapted in each case
to some specific 3+1 evolution formalism:
Fritelli-Reula~\cite{Stewart}, KST~\cite{Calabrese03,ST05},
BSSN-NOR~\cite{GMG04}, or Z4~\cite{CP Z4}. The focus changed
suddenly after 2005 by the impact of a breakthrough: the first
'\,long term' binary-black-hole simulation, which was achieved in
a generalized-harmonic formalism~\cite{Pretorius05}. A series of
constraint-preserving boundary conditions proposals in this
framework started then~\cite{Babiuc07,Kreiss07}, and continues
today~\cite{Winicour,Rinne}.

We will retake in this paper the 3+1 approach to
constraint-preserving boundary conditions, following the way
opened very recently for the BSSN case~\cite{Nunez}. More
specifically, we will revisit the Z4 case, not just because of its
intrinsic relevance, but also for its relationship with other 3+1
formulations (BSSN, KST, see refs.~\cite{Z48,Z4BSSN} for details).
Also, the close relationship between the Z4 and the
generalized-harmonic formulations suggest that our results could
provide a different perspective in this other context. This was
actually what happened with the current constraint-damping terms:
first derived in the Z4 context~\cite{damping} and then applied
successfully in generalized-harmonic
simulations~\cite{Pretorius05}.

Our results are both at the theoretical and the numerical level.
In section II, we consider the first-order Z4 formalism in normal
coordinates (zero shift) for the harmonic slicing case. This case
was known to be symmetric-hyperbolic for a particular choice of
the parameter which controls the ordering of space
derivatives~\cite{Z4,LNP}. We extend this result to a range of
this ordering parameter, by providing explicitly a
positive-definite energy estimate. Then we use this estimate for
deriving algebraic constraint-preserving boundary conditions both
for the energy and the normal momentum components.

In section III we consider the dynamical evolution of constraint
violations (subsidiary system). Following standard
methods~\cite{Calabrese03,ST05,GMG04}, we transform algebraic
boundary conditions of the subsidiary system into derivative
boundary conditions for the main system. We introduce a new basis
of dynamical fields in order to revise the constraint-preserving
conditions proposed in refs.~\cite{CP Z4, LNP} for the Z4
formalism, including also a new coupling parameter which affects
the propagation speeds of the (modified) incoming modes. In the
case of the energy constraint, we get a closed subsystem for the
principal part, allowing an analytical stability study at the
continuum level which is presented in Appendix B.

A simple numerical implementation of the proposed conditions is
given in section IV, where we test the stability in the linear
regime, by considering small random-noise perturbations aroun flat
space (robust stability test). The results show the numerical
stability of the proposed boundary conditions in this regime for
many different combinations of the parameters. The space
discretization scheme is the simplest one with the
summation-by-parts (SBP) property~\cite{Olsson}. In this way we
avoid masking the effect of our conditions (at the continuum
level) with the effect of more advanced space-discretization
algorithms, like FDOC~\cite{JCPpaper} devised to reduce the high
frequency noise level in long-term simulations, which has recently
been applied to the black-hole case~\cite{polyvalent}. For a
comparison, we run also with periodic boundary conditions, where
the noise level keeps constant. The proposed boundary conditions
produce instead a very effective decreasing of (the cumulated
effect of) energy and momentum constraint violations. In the case
of cartesian-like grids, we also compare the standard '\,a la
Olsson' treatment~\cite{Olsson}, with a modified numerical
implementation which does not use the corner and vertex points,
avoiding in this way some stability issues and providing much
cleaner evidence of constraint preservation.

In section V, we test the non-linear regime with the Gowdy
waves~\cite{Gowdy71} metric, one of the standard numerical
relativity code tests, as we have done recently for the energy
constraint case~\cite{bilbao}. We endorse in this way some recent
claims (by Winicour and others) that the current code
cross-comparison efforts~\cite{Mexico,Mexico2} should be extended
to the boundaries treatment. A convergence test is performed
against this exact strong-field solution, showing the expected
convergence rate (second order for our simple SBP method). Testing
the proposed boundary conditions results into a stable and
constraint-preserving behavior, in the sense that energy and
momentum constraint violations remain similar or even smaller than
the corresponding effects with exact (periodic or reflection)
boundary conditions for the Gowdy metric.

\section{The Z4 case revisited}

We will consider here the Z4 evolution system:
\begin{equation}\label{Z4}
  R_{\mu \nu} + \nabla_{\mu} Z_{\nu} + \nabla_{\nu} Z_{\mu} =
  8\; \pi\; (T_{\mu \nu} - \frac{1}{2}\;T\; g_{\mu \nu}).
\end{equation}
More specifically, we will consider the first-order version in
normal coordinates, as described in refs.~\cite{Z48,LNP}. For
further convenience, we will recombine the basic first-order
fields $(K_{ij},\,D_{ijk},\,A_i,\,\Theta,\,Z_i)$ in the following
way:
\begin{eqnarray}
\label{Pimu a}
  \Pi_{ij} &=& K_{ij} - (\,tr K - \Theta)\, \gamma_{ij} \\
\label{Pimu b}  V_i &=& \gamma^{rs}(D_{irs} - D_{ris}) - Z_k \\
\label{Pimu c}  \mu_{ijk} &=& D_{ijk} - (\gamma^{rs}D_{irs}- V_i)\,\gamma_{jk}\\
\label{Pimu d}  W_i &=& A_i - \gamma^{rs}D_{irs} + 2\,V_i
\end{eqnarray}
so that the new basis is
$(\Pi_{ij},\,\mu_{ijk},\,W_i,\,\Theta,\,V_i)$. Note that the
vector $Z_i$ can be recovered easily as
\begin{equation}\label{Zfrom mu}
    Z_i=-{\mu^k}_{ik}.
\end{equation}

With this new choice of basic dynamical fields, the principal part
of the evolution system gets a very simple form in the harmonic
slicing case:
\begin{eqnarray}
\label{evol a}  \partial_t\,W_i &=& \cdots \\
\label{evol b}  \partial_t\,\Theta + \alpha~\partial_k\,V^k &=& \cdots \\
\label{evol c}  \partial_t\,V_i + \alpha~\partial_i\, \Theta&=& \cdots \\
\label{evol d}  \partial_t\,\Pi_{ij} + \alpha~\partial_k \,{\lambda^k}_{ij}&=& \cdots \\
\label{evol e} \partial_t \,\mu_{kij} + \alpha~\partial_k
\,\Pi_{ij} &=& \cdots
\end{eqnarray}
where the dots stand for non-principal contributions, and we have
noted for short
\begin{equation}\label{lambda}
    \lambda_{kij} = \mu_{kij} + \gamma_{k(i}W_{j)} - W_k\,\gamma_{ij}
    - (1+\zeta)\,[\,\mu_{(ij)k}+\gamma_{k(i}Z_{j)}\,]\,,
\end{equation}
where $~\zeta~$ is a space-derivatives ordering parameter and
round brackets denote index symmetrization.

The first-order version of the Z4 system is known to be
symmetric-hyperbolic in normal coordinates with harmonic slicing,
at least for the usual ordering $~\zeta=-1~$~\cite{Z48}. It
follows from (\ref{evol a}-\ref{evol e}) that this result can be
extended to the following range of the ordering parameter
\begin{equation}\label{range}
    -1 \le \zeta \le 0\,,
\end{equation}
which covers the symmetric ordering case ($\zeta=0$). The
corresponding 'symmetrizer', or 'energy estimate', can be written
as:
\begin{equation}\label{Eestimate}
    S = \Theta^2 + V_kV^k + \Pi^{ij}\,\Pi_{ij}
    + \tilde{\mu}^{kij}\tilde{\mu}_{kij}
    + (1+\zeta) (Z^kZ_k-\tilde{\mu}^{kij}\tilde{\mu}_{ijk})
     + 2\, \zeta\, Z_kW^k\,,
\end{equation}
where we have noted for short
\begin{equation}\label{mutilde}
    \tilde{\mu}_{kij} = \mu_{kij} - W_k\, \gamma_{ij}\,.
\end{equation}
Allowing for (\ref{evol a}-\ref{evol e}), we get
\begin{equation}\label{div term}
    \frac{-1}{2\,\alpha}~\partial_t\,S =
    \partial_k\,(\Theta\, V^k + \Pi_{ij}\, \lambda^{kij}) + \cdots
\end{equation}
and the divergence theorem can be used in order to complete the
proof. The positivity proof for $~S~$ for the interval
(\ref{range}) is given in Appendix A.

\subsection*{Characteristic decomposition}

We can consider now some specific space surface, in order to
identify the constraint modes by looking at the evolution
equations for $\Theta$ and $Z_i$ in the system (\ref{evol
a}-\ref{evol e}). It follows from (\ref{evol b}, \ref{evol c})
that the energy-constraint modes are given by the pair
\begin{equation}\label{Energy pair}
     E^\pm = \Theta~\pm~V_n
\end{equation}
with propagation speed $\pm\alpha$ (the index $n$ meaning the
projection along the unit normal $n_i$). Also, allowing for
(\ref{Zfrom mu},\ref{evol e}), we can easily recover the evolution
equation for $Z_i$, namely
\begin{equation}\label{evol Z}
    \partial_t \,Z_{i} - \alpha~\partial_k \,{\Pi^k}_{i} = \cdots
\end{equation}
so that we can identify the momentum-constraint modes with the
three pairs, with propagation speed $\pm\alpha$,
\begin{equation}\label{Momentum pairs}
     M_i^\pm = \Pi_{ni}~\pm~\lambda_{nni}~.
\end{equation}
Note that, allowing for (\ref{Pimu a}), the normal component
$\Pi_{nn}$ does correspond with the transverse-trace component of
the extrinsic curvature $K_{ij}$. We give for completeness the
remaining modes, the fully tangent ones, with propagation speed
$\pm\alpha$,
\begin{equation}\label{Transverse pairs}
     T_{AB}^\pm =\Pi_{AB}~\pm~\lambda_{nAB}
\end{equation}
(capital indices denote a projection tangent to the surface), and
the standing modes (zero propagation speed):
\begin{equation}\label{standing}
    W_i~,\qquad V_A ~,\qquad \mu_{Aij}~.
\end{equation}

\subsection*{Algebraic boundary conditions}

We can take advantage of the positive-definite energy estimate
(\ref{Eestimate}) in order to derive suitable algebraic boundary
conditions. We can integrate (\ref{div term}) in space and, by
applying the divergence theorem, we get a positivity condition for
the boundary terms, namely
\begin{equation}\label{bound term}
    (\Pi^{ij}~\lambda_{nij} + \Theta~V_n)\mid_\Sigma ~\ge~ 0
\end{equation}
where $\Sigma$ stands for the boundary surface ($\mathbf{n}$ being
here its outward normal).

The contribution of the fully tangent modes (\ref{Transverse
pairs}), independent of the energy and momentum sectors, is given
by
\begin{equation}\label{bound transv}
    \Pi^{AB}~\lambda_{nAB} = \frac{1}{4}~tr\,[(T^+)^2-(T^-)^2]\,,
\end{equation}
so that the contribution of these modes to the boundary term in
(\ref{bound term}) will be non-negative if we impose the standard
algebraic boundary-conditions:
\begin{equation}\label{transv algeb}
    T^-_{AB} = \sigma\,T^+_{AB}\qquad |\sigma|\le 1\,,
\end{equation}
the case $\sigma=0$ corresponding to maximal dissipation. A less
strict condition is obtained by adding an inhomogeneous term,
namely
\begin{equation}\label{transv algeb bis}
    T^-_{AB} = \sigma\,T^+_{AB} + G_{AB}\,.
\end{equation}
This can cause some growth of the energy estimate but, provided
that the array $G$ consists of prescribed spacetime functions, the
growth rate can be bounded in a suitable way so that a well-posed
system can still be obtained (see for instance
refs.~\cite{GMG04,Nunez}).

This simple strategy, when applied to the energy and momentum
modes (\ref{Energy pair}, \ref{Momentum pairs}) is not compatible
with constraint preservation in the generic case (see also
ref.~\cite{ST05}). For the energy sector, constraint preservation
is obtained only for the extreme case:
\begin{equation}
    E^- = E^+\qquad \Leftrightarrow\qquad \Theta=0\,,
\end{equation}
which will reflect energy-constraint violations back into the
evolution domain. These conditions would be then of a limited
practical use in realistic simulations.

A different approach can be obtained by realizing that the
contribution to the boundary term in (\ref{bound term}) would have
the right sign if one uses the following '\,logical gate'
condition:
\begin{equation}\label{Thetagate}
    \Theta\mid_\Sigma ~ = 0\qquad
    if\qquad (\Theta~V_n)\mid_\Sigma ~< ~ 0
\end{equation}
($\Theta$-gate in ref.~\cite{bilbao}). It is clear that the
boundary condition (\ref{Thetagate}) preserves the energy
constraint, as it modifies just the $\Theta$ values, by setting
them to zero when the condition is fulfilled, without affecting
any other dynamical field.

The same strategy can work for normal components of the momentum
modes (\ref{Momentum pairs}), at least for the symmetric choice of
the ordering parameter. Allowing for (\ref{lambda}), one has
\begin{equation}\label{Znmode}
    M^\pm_n = \Pi_{nn} \mp Z_n\qquad (\zeta=0)\,,
\end{equation}
so that a constraint-preserving (reflection) condition can be
obtained in the extreme case as well. In the logical gate
approach, the contribution of the modes (\ref{Znmode}) to the
boundary term in (\ref{bound term}) will have the right sign if
one uses the condition (case $\zeta=0$ only):
\begin{equation}\label{Zngate}
    Z_n\mid_\Sigma ~ = 0\qquad
    if\qquad (Z_n~\Pi_{nn})\mid_\Sigma ~> ~ 0\,,
\end{equation}
which clearly preserves the normal component of the momentum
constraint.

For the tangent momentum modes $M^\pm_A$ (tangent to the boundary
surface), however, the contribution in (\ref{bound term}) will be
\begin{equation}\label{bound mix}
   2\,\Pi^{nA}~\lambda_{nnA}\,,
\end{equation}
where $\lambda_{nnA}$ is inhomogeneous in $Z_A$ for any value of
the ordering parameter. Moreover, the inhomogeneous terms are not
prescribed functions, but rather some combinations of dynamical
fields. A different strategy must then be devised in this case, as
we will see below.

\section{Constraints evolution and derivative boundary conditions}

The time evolution of the energy-momentum constraints can be
easily derived by taking the divergence of the Z4 field equations
(\ref{Z4}), that is
\begin{equation}\label{WaveZ}
  \Box~Z_{\mu} + R_{\mu\nu} Z^\nu = 0~.
\end{equation}
We can write down the second order equation (\ref{WaveZ}) as a
first order system and impose then maximally dissipative boundary
conditions on (the first derivatives of) the $Z_{\mu}$ components.
In this way, the boundaries will behave as one-way membranes for
constraint-violating modes, at least for the ones propagating
along the normal direction $n_i$.

The procedure can be illustrated with the energy-constraint, that
is the time component of (\ref{WaveZ}):
\begin{equation}\label{Thetasub}
\partial^2_{tt}~\Theta - \alpha^2\,\triangle~\Theta = \cdots
\end{equation}
A first-order version can be obtained as usual by considering
first-order derivatives as independent quantities, namely
\begin{equation}
    \dot{\Theta} \equiv 1/\alpha~\partial_t\,\Theta\,, \qquad
    \Theta_k \equiv \partial_k\,\Theta\,.
\end{equation}
We can write then (\ref{Thetasub}) as the following first-order
symmetric-hyperbolic system
\begin{eqnarray}
\label{Thetasubfirst}
\partial_{t}~\dot{\Theta} - \alpha~\partial_k\,\Theta^k &=&
\cdots\\
\partial_{t}~\Theta_k - \alpha~\partial_k\,\dot{\Theta} &=& \cdots
\end{eqnarray}

Boundary conditions for (the incoming modes of) the subsidiary
system can be enforced then in the standard way. We will consider
here for simplicity the 'maximal dissipation' condition, that is
(we assume that the boundary is on the right):
\begin{eqnarray}
  \dot{\Theta} &+& n^k\Theta_k = 0\,.
  \label{Thetabound}
\end{eqnarray}
Now we can use it as a tool for setting up boundary conditions for
the energy modes of the main evolution system. One can for
instance enforce directly (\ref{Thetabound}), as in
ref.~\cite{bilbao}.

We will rather use (\ref{Thetabound}) as a tool for getting
(derivative) boundary conditions for the incoming energy mode of
the evolution system (\ref{evol a} - \ref{evol e}). To do this, we
can use the evolution equation (\ref{evol b}) for transforming
(\ref{Thetabound}) into a convenient version of the energy
constraint, namely:
\begin{eqnarray}\label{Energy constraint}
  {\cal E} &=& \partial_k\,V^k - \partial_n~\Theta + \cdots
\end{eqnarray}
We can now use (\ref{Energy constraint}) in order to modify the
evolution equation of the incoming energy mode $E^-$, that is:
\begin{eqnarray}\label{E- modif}
  1/\alpha~\partial_t~E^- + \partial_k\,V^k - \partial_n\,\Theta &=&
  a~{\cal E} + \cdots
\end{eqnarray}
The whole process is equivalent to the simple replacement:
\begin{eqnarray}\label{E- repl}
  E^- &\rightarrow& E^- + a~(\Theta^{(adv)} -\Theta)
\end{eqnarray} where $\Theta^{(adv)}$ is the solution of the
advection equation (\ref{Thetabound}).

The choice $a=1$ corresponds to the standard
recipe~\cite{Calabrese03,ST05,GMG04} of '\,trading' space normal
derivatives by time derivatives, in the incoming modes. This
implies that the modified mode gets zero propagation speed along
the given direction $\mathbf{n}\,$. In this case, allowing for
(\ref{E- modif}), the time derivative of $E^-$ would actually
vanish, modulo non-principal terms; this amounts to freezing  the
incoming modes to their initial values (maximal dissipation '\,on
the right-hand-side'), which is a current practice in some
Numerical Relativity codes. Note however that constraint
preservation requires using the right non-principal terms, that
can be deduced from the full expression (\ref{E- repl}).

The choice $a=2\,$ would imply instead that the modified mode gets
the same positive speed ($+\alpha$) than the outgoing one $E^+$.
We show in Appendix B that this choice will lead to a
weakly-hyperbolic (ill-posed) boundary system. Our results confirm
that $a=1$ is actually a safe
choice~\cite{Calabrese03,ST05,GMG04}, although other values in the
interval $~1\le\,a<2~$ lead also to a strongly hyperbolic system
with non-negative speeds for all energy modes (see Appendix B for
details).

\subsection*{Momentum constraint conditions}
The same method can be applied to the momentum constraint modes,
although in a less straightforward way. Let us start from the
evolution equation (\ref{evol Z}) for $Z_i$, and take one extra
time derivative. We get in this way
\begin{eqnarray}
\partial^2_{tt}~Z_i + \alpha^2\,\partial^2_{rs}~{\lambda^{rs}}_i
 &=& \cdots
 \label{Zsub}
\end{eqnarray}
which, after some cross-derivatives cancellations, leads to the
space components of (the principal part of) the covariant equation
(\ref{WaveZ}).

A first-order version of (\ref{Zsub}) can be obtained again by
considering first-order derivatives as independent quantities. For
the time derivative we will take the obvious choice
\begin{equation}
    \dot{Z_i} \equiv 1/\alpha~\partial_t\,Z_i\,.
\end{equation}
The treatment of space derivatives, however, is complicated by the
fact that we are dealing with a first-order formulation, so that
there are additional ordering constraints to be allowed for.
Following refs.~\cite{CP Z4,LNP}, we will define for further
convenience
\begin{equation}\label{Zkiexplicit}
  Z_{ki} \equiv \partial_k\,Z_i
 - \partial_{[\,k}\,A_{i]}- \partial_{[\,k}\,D_{i]}
 - (1-\zeta)\,\gamma^{rs}\,\partial_{[\,r}\,D_{k]\,is}
 + (1+\zeta)\,\gamma^{rs}\,\partial_{[\,r}\,D_{i]\,ks}\,,
\end{equation}
where we have noted for short $D_i = \gamma^{rs}D_{irs}$\,. A
closer look to (\ref{Zkiexplicit}) shows that $Z_{ki}$ is just the
space derivative of $Z_i$, modulo ordering constraints. In the
notation of this paper:
\begin{equation}\label{Zki}
    Z_{ki} = -\partial_r\,{\mu_{ki}}^r + \partial_{[i}\,W_{k]}
    + (1+\zeta)~[\,\partial_r\,{\mu_{(ki)}}^r+\partial_{(k}\, Z_{i)}\,]
    + \cdots
\end{equation}

We can write now (\ref{Zsub}) in the first-order form
\begin{eqnarray}
\partial_{t}~\dot{Z_i} - \alpha~\partial_{k}~{Z^{k}}_i
 &=& \cdots
 \label{Zsubfirst}\\
\partial_{t}~{Z_{ki}} - \alpha~\partial_{k}~\dot{Z_i} &=& \cdots
\end{eqnarray}
which is a symmetric-hyperbolic first-order version of the
momentum-constraint evolution system (other versions could be
obtained by playing with the ordering constraints in a different
way). The vanishing of the incoming modes of this subsidiary
system can be enforced now in the same way as for the energy
constraint, namely:
\begin{eqnarray}
  \dot{Z_i} &+& n^kZ_{ki}= 0\,.
  \label{Zbound}
\end{eqnarray}
This is obviously a 'maximal dissipation' constraint-preserving
condition for the subsidiary system, which can be used for to get
a derivative boundary condition for the main evolution system, as
we did for the energy modes in the preceding subsection. To be
more specific, we can use the evolution equation (\ref{evol Z})
for transforming (\ref{Zbound}) into a convenient version of the
momentum constraint, that is
\begin{eqnarray}
  {\cal M}_i &=& \partial_k\,{\Pi^k}_i + n^kZ_{ki}
   + \cdots \label{Momentum constraint}
\end{eqnarray}
and use it for modifying the evolution equation of the incoming
momentum modes $M^-_i$, namely:
\begin{eqnarray}
  1/\alpha~\partial_t~M^-_i + \partial_k\,{\lambda^k}_{ni}
  -  \partial_n\,\Pi_{ni}&=&-a~{\cal M}_i + \cdots  \label{M- modif}
\end{eqnarray}
which amounts to the following replacement:
\begin{eqnarray}
  M^-_i &\rightarrow& M^-_i + a~(Z^{(adv)}_i - Z_i)\,,
\label{M- repl}
\end{eqnarray}
where $Z_i^{(adv)}$ is the solution of the advection-like equation
(\ref{Zbound}).

The choice $a=1$ would imply again that the modified modes get
zero propagation speeds along the normal direction, whereas the
choice $a=2$ would imply instead that the modified modes get the
same positive speed ($+\alpha$) than the outgoing ones $M^+_i$.
This result requires the extra ordering terms in
(\ref{Zkiexplicit}): this was actually the reason for including
them. Note that we can consider different values of the coupling
parameter for the energy modes ($\,a=a_E\,$), and even for the
normal and tangent momentum modes ($\,a=a_N\,,~a_T\,$,
respectively).

For any value $a \ge 1$, the modified modes can be computed
consistently from inside. The momentum system however is too
complicated for a full hyperbolicity analysis, like the one we
provide in Appendix B for the energy sector. Part of the
complication comes from the coupling with the non-constraint
modes, which require their own boundary conditions. Let us
remember at this point that the boundary conditions presented in
this section are derivative, not algebraic. This means that, even
in the symmetric hyperbolic cases, proving well-posedness is by no
means trivial.

For that reason, we will rather follow the approach of
ref.~\cite{ST05}, focusing in the stability of small perturbations
around smooth solutions, which can be tested numerically. We start
in the following section, by performing a '\,robust stability'
test in order to check the numerical stability of high-frequency
perturbations around the Minkowsky metric. As a full set of
boundary conditions is required, even in this weak-field test, we
supplement our conditions for the constraint-related modes with
the freezing of the initial values of the incoming non-constraint
modes (maximal dissipation 'on the right-hand-side').

\section{Numerical implementation}

Let us test now the stability and performance of the proposed
conditions in the linear weak-field regime, by considering a small
perturbation of Minkowski space-time, which is generated by taking
random data both in the extrinsic curvature and in the
constraint-violation quantities $(\Theta,\,Z_i)$. In this way the
initial data violate the energy-momentum constraints, but preserve
all ordering constraints. The level of the random noise will be of
the order $10^{-6}$, small enough to make sure that we will keep
in the linear regime during the whole simulation (Robust Stability
test, see ref.~\cite{Mexico} for details).

We will use the standard method of lines~\cite{MoL} as a finite
difference algorithm, so that space and time discretization will
be treated separately. The time evolution will be dealt with a
third-order Runge-Kutta algorithm. The time step ${\rm d}t$ is
kept small enough to avoid an excess of numerical dissipation that
could distort our results in long runs.

For space discretization, we will consider a three-dimensional
rectangular grid, evenly-spaced along every space direction, with
a space resolution $h=1/80$. We will use there the simplest
centered, second-order-accurate, discretization scheme. At the
points next to the boundary, where we can not use the required
three-points stencil, we will switch to the standard first-order
upwind (outgoing) scheme. This combination is the simplest one
with the summation-by-parts (SBP) property~\cite{Olsson}. In this
way we expect that the theoretical properties derived from
symmetric-hyperbolicity will show up in the simulations in a more
transparent way.  For the same reason, we avoid adding extra
viscosity terms that could mask the effect of our conditions (at
the continuum level) with the dissipative effects of the
discretization algorithm. Just to make sure, we run also with
periodic boundary conditions, where the noise level keeps
constant: any decrease of the constraint-violation level will then
be due to the proposed conditions, not to the discretization
scheme.

Let us be more specific about the boundary treatment. At boundary
points, we use the first-order upwind algorithm in order to get a
prediction for every dynamical field. Once we have got this
prediction, we perform the characteristic decomposition along the
direction normal to the boundary. The predicted values for the
outgoing modes, for which the upwind algorithm is known to be
stable, will be kept (this includes the 'standing' modes, with
zero characteristic speed). The (unstable) incoming modes will be
replaced instead by the values arising from our boundary
conditions, as described in the preceding section.

We start with simulations in which the proposed conditions are
applied just to the $z$ face, whereas we keep periodic boundary
conditions along the $x$ and $y$ directions. In this way we can
detect instabilities which are inherent to the proposed boundary
conditions on smooth boundaries (no corners), allowing at the same
time for some non-trivial dynamics along at least one tangent
direction.

\begin{figure}[h]
\centering
\includegraphics[width=10cm, height=6cm]{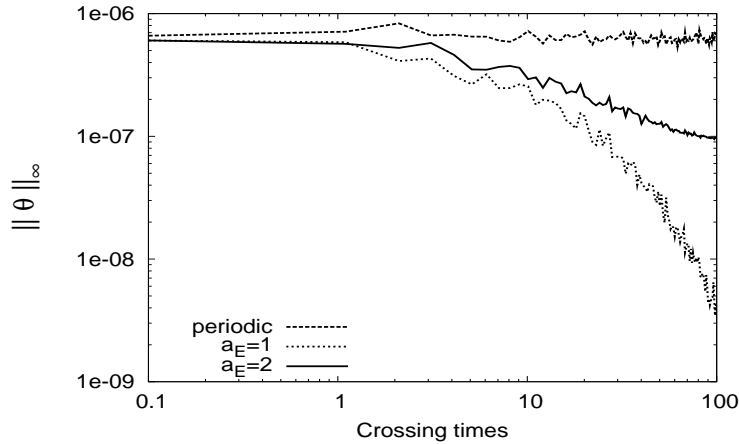}
\caption{Robust stability test. Time evolution of the maximum norm
of $\Theta$, with just one face open (periodic boundaries are
implemented along the transverse directions). The fully periodic
boundaries result (dashed lines) is also included for comparison.
We see some growing mode onset in the $a_E=2$ case, whereas the
constraint-preserving $a_E=1$ case (continuous line) is very
efficient at reducing the initial noise level.}\label{robustTheta}
\end{figure}

We plot in fig.~\ref{robustTheta} the maximum norm of the
energy-constraint-violating quantity $\Theta$ for two different
choices of the coupling parameter of the energy mode:
$a_E=1\,,2\,$. We can see that, after $100$ crossing times, the
case $a_E = 2$ starts showing the effect of the linear modes
predicted by our hyperbolicity analysis in Appendix B, by
departing from the maximal dissipation pattern of decay. We plot
for comparison the results obtained by applying periodic boundary
conditions, so we can see how, for the choice $a_E = 1$, the
proposed constraint-preserving conditions are extremely effective
at 'draining out' energy constraint violations. The rate of decay
is actually the same as the one obtained by applying maximal
dissipation conditions '\,on the right-hand-side' also to the
energy modes, as expected from the analysis given in the previous
section. In what follows, we will fix $a_E = 1$ for this coupling
parameter.

\begin{figure}[b]
\centering
\includegraphics[width=8cm, height=6cm]{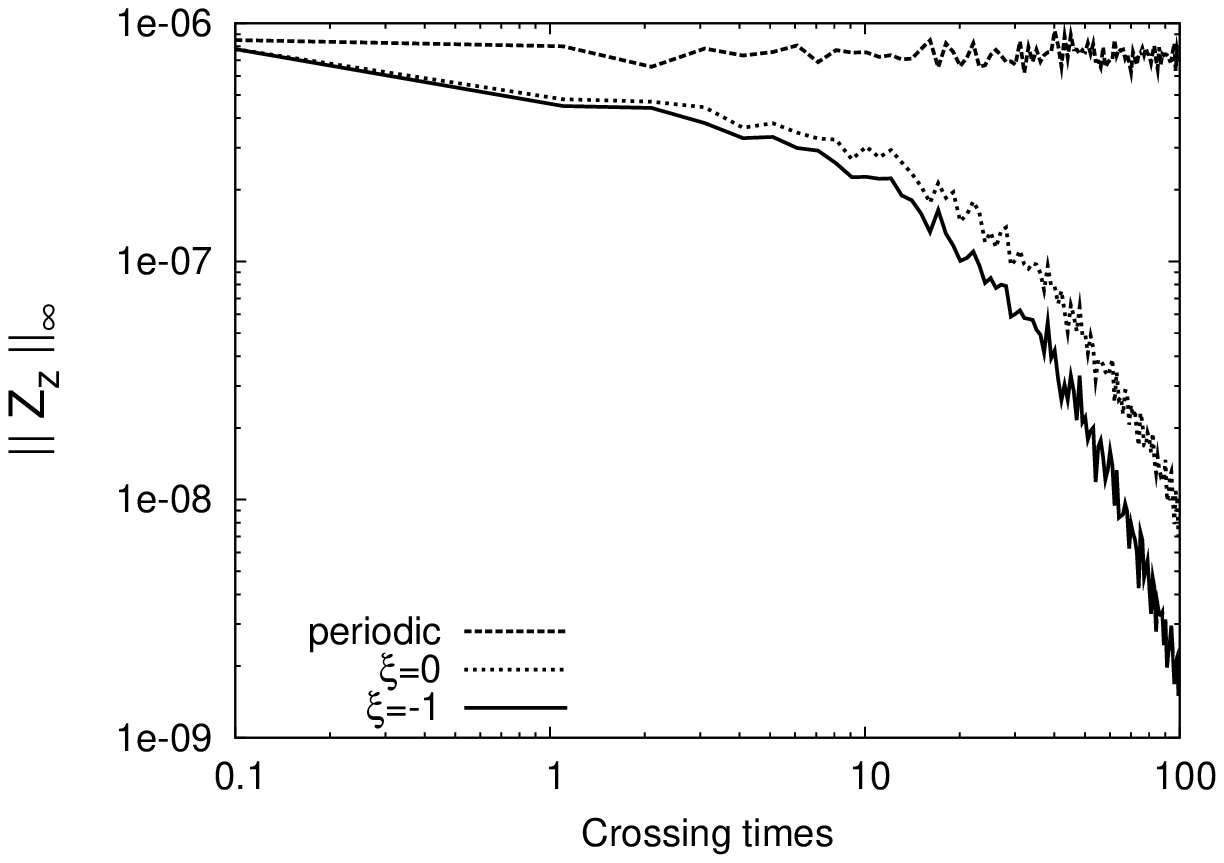}
\includegraphics[width=8cm, height=6cm]{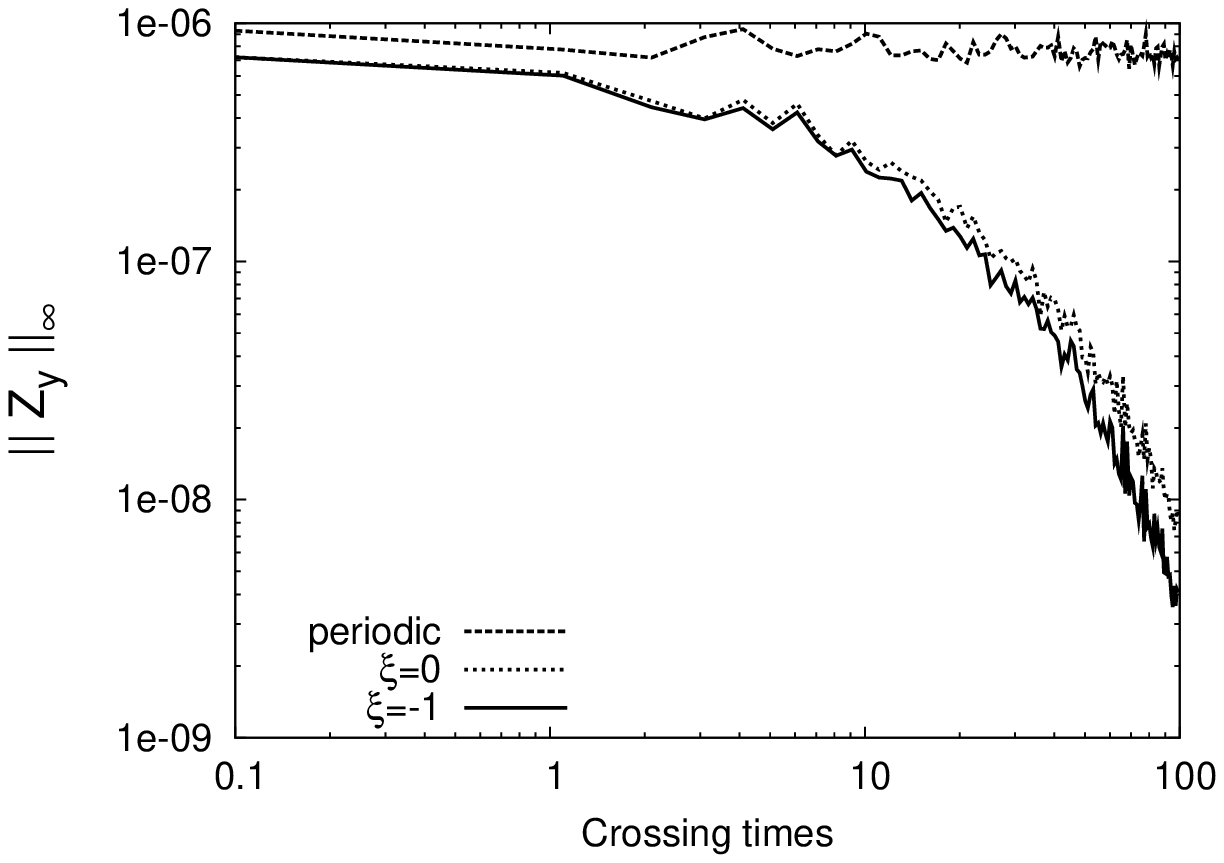}
\caption{Robust stability test. Time evolution of the maximum norm
of the longitudinal and transverse $Z_i$ components (left and
right panels, respectively). In both cases, the periodic
boundaries results (dashed lines) are included for comparison. The
initial noise in the momentum constraint gets reduced very
efficiently in both the $\zeta=-1$ and the $\zeta=0$ cases,
although there is a slight difference, more visible in the
longitudinal case (left panel).}\label{robust1}
\end{figure}

We plot in fig.~\ref{robust1} both the maximum norm of the
longitudinal (left panel) and transverse components (right panel)
of the momentum-constraint-violating vector $Z_i$ for the choice
$\,a_N=a_{\,T} = 1\,$ of the coupling parameter of the momentum
modes. We include again for comparison the results obtained by
applying periodic boundary conditions, so we can see how the
proposed constraint-preserving conditions are very effective at
'draining out' energy constraint violations. The $Z_y$ plots are
slightly, sensitive to the ordering parameter $\zeta=0,\,-1$. In
the $\zeta=-1$ case, the rate of decay is actually the same as the
one obtained by applying instead maximal dissipation conditions
'\,on the right-hand.-side' for the momentum modes. The results
are qualitatively the same for other components of $Z_i$ and for
other parameter combinations $\,a_N,\,a_{\,T} = 1,\,2\,$.

\begin{figure}[h]
\centering
\includegraphics[width=8cm, height=6cm]{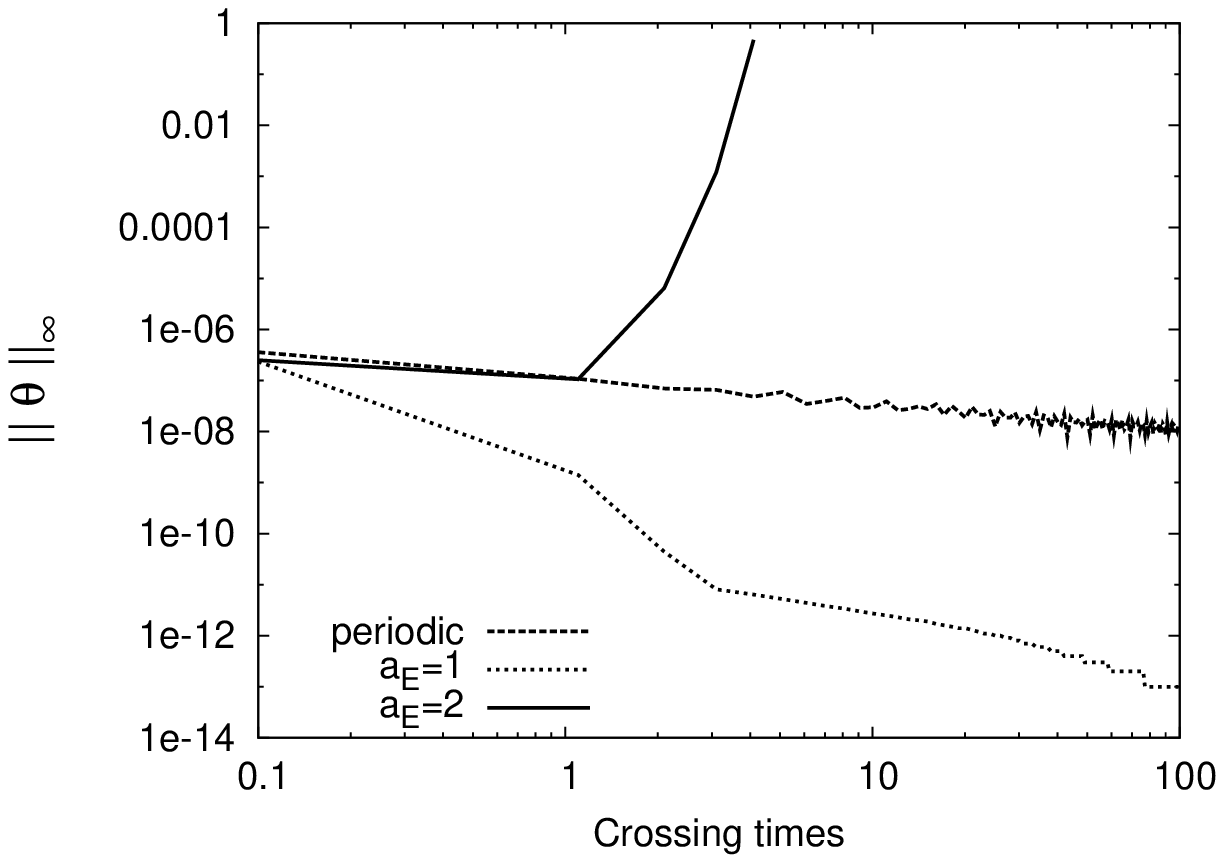}
\includegraphics[width=8cm, height=6cm]{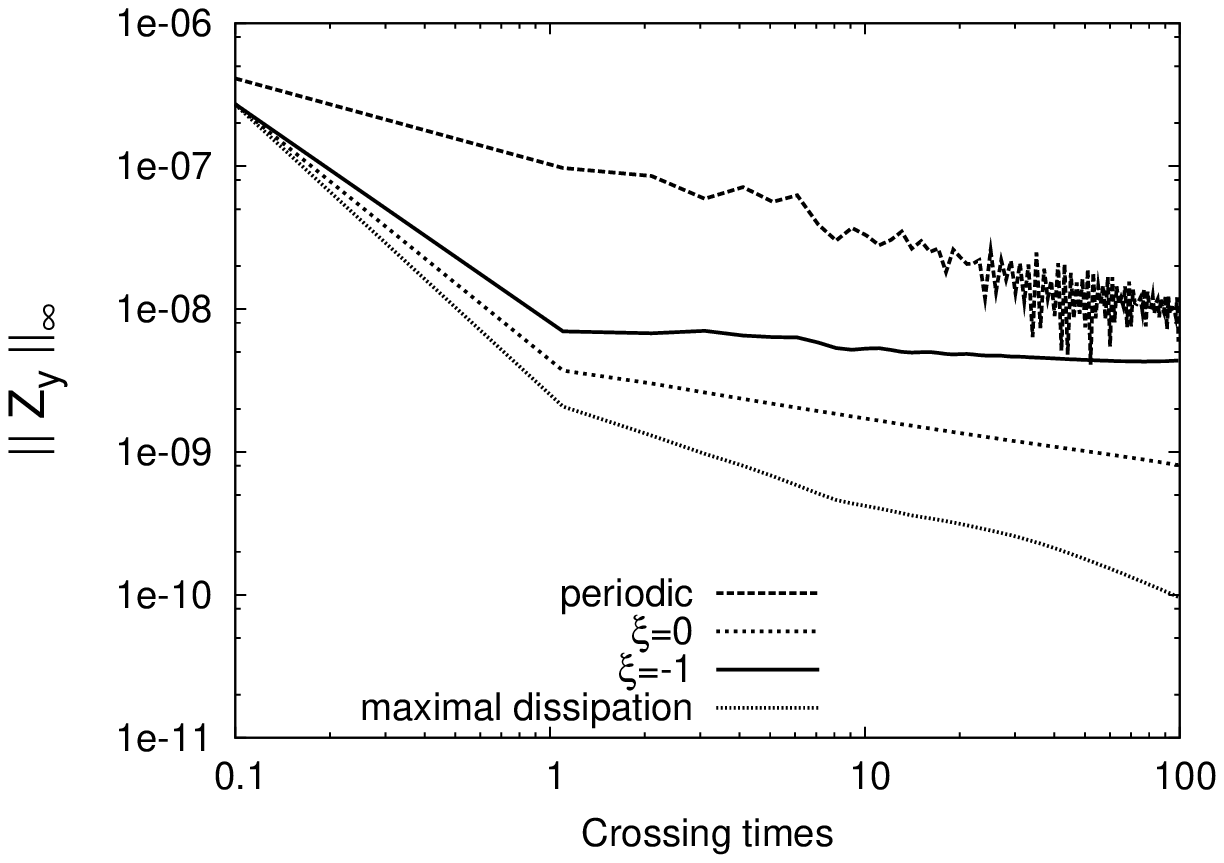}
\caption{Robust stability test. Time evolution of the maximum norm
of the constraint-violating quantities $\Theta$ (left panel), and
$Z_y$ (right panel). The proposed boundary conditions are applied
here to all faces, including corners and vertices. Some amount of
numerical dissipation has been added, so that the periodic
boundaries plots (dashed lines) get a visible negative slope. The
choice $a_E=1$ for the energy modes is still clearly stable (left
panel). The choice $\zeta=-1$ for the momentum modes (right panel)
shows a growing mode onset. For comparison, a plot with the
maximal dissipation results is also included in the right panel
(bottom line).}\label{robustOlsson}
\end{figure}

In order to perform a full test for cartesian-like grids,
including corner and vertex points, we will repeat the same
simulations, but this time with the proposed boundary conditions
applied to all faces, not just to the $z$ ones. A standard
treatment of corner points '\,a la Olsson'~\cite{Olsson}, like the
one presented in previous works~\cite{CP Z4, LNP}, results into
numerical instability issues. A simple cure is to add some extra
dissipation at the interior points, at the price of masking the
theoretical results, et the continuum level, with the numerical
viscosity effects, as shown in fig.~{\ref{robustOlsson}}. We can
see there that opening all faces makes the effects to appear much
faster. The expected instability of the $a_E=2$ choice of the
energy coupling parameter, which was just an onset in
fig.~\ref{robustTheta}, shows up manifestly here (left panel).
Also, a growing mode onset is clearly visible for the choice
$\zeta=-1$ of the momentum-constraint coupling parameter (right
panel). The case $\zeta=0$ looks stable, although no strong
conclusion can be drawn because of the added numerical
dissipation. Maximal dissipation results are also shown for
comparison in the right panel (bottom line).

\begin{figure}[h]
\centering \scalebox{0.3}[0.3] {\includegraphics{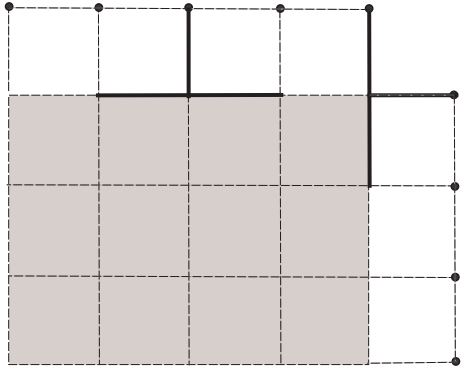}}
\caption{Stencil for first-order prediction at boundary points
(black dots). Interior points belong to the shaded zone. The
stencil for two boundary points is represented by thick lines (one
space direction has been suppressed for clarity). Note that no
corner points are required, as the tangent derivatives are not
computed at the boundary, but at the neighbor
layer.}\label{stencil}
\end{figure}

We will present here an alternative numerical treatment. At
boundary points, tangent derivatives are computed at the
next-to-last layer. The corresponding stencil is shown in
fig.~{\ref{stencil}}. In this way the corner points are not
required. This avoids the reported code stability issues, even
without adding extra numerical dissipation terms. Note that
transverse derivatives are still computed using the standard
three-point SBP algorithm, like in the smooth boundaries case. As
every space derivative can be considered separately (we are
dealing with a first-order system) the SBP property should still
follow for our modified scheme. The price for the shift of the
transverse derivatives to the next-to-last layer is getting just
first-order accuracy at the boundary, but the longitudinal
derivatives there were yet only first-order accurate anyway.

\begin{figure}[h]
\centering
\includegraphics[width=8cm, height=6cm]{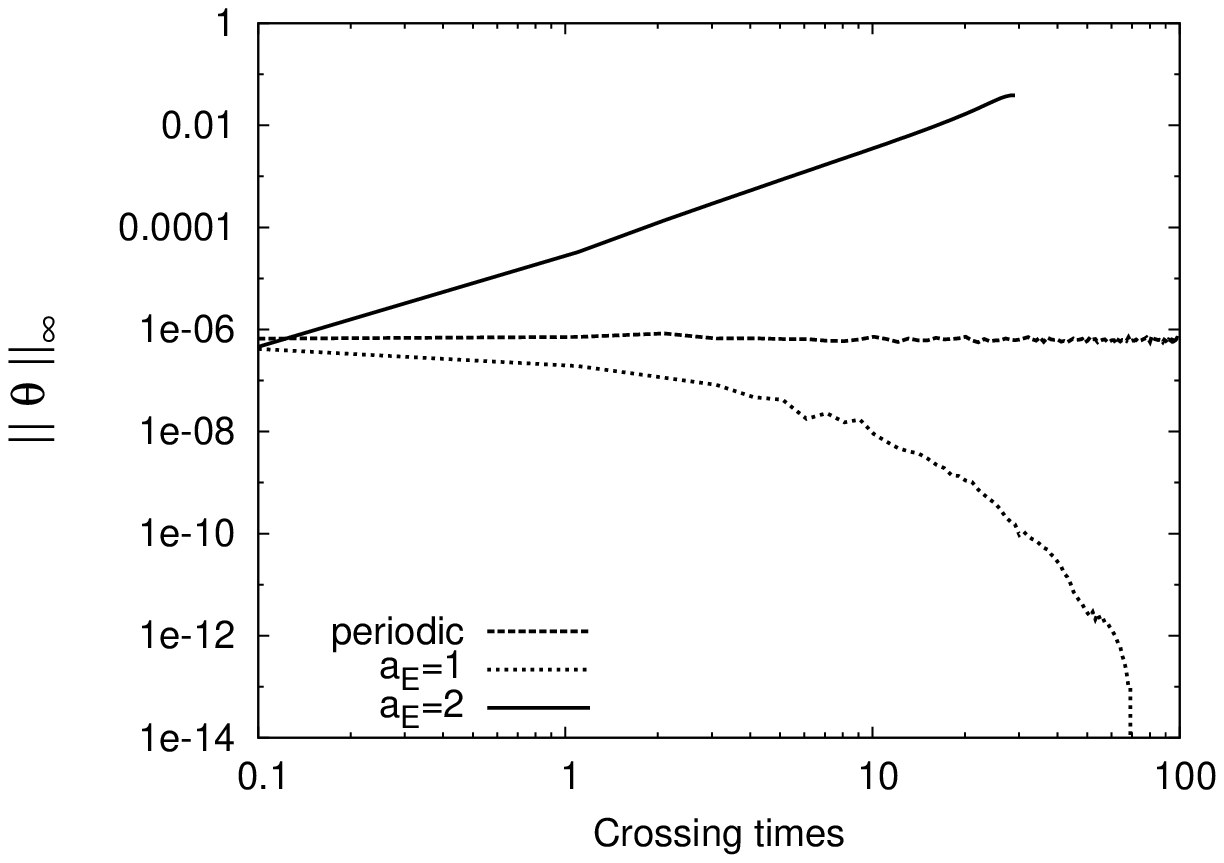}
\includegraphics[width=8cm, height=6cm]{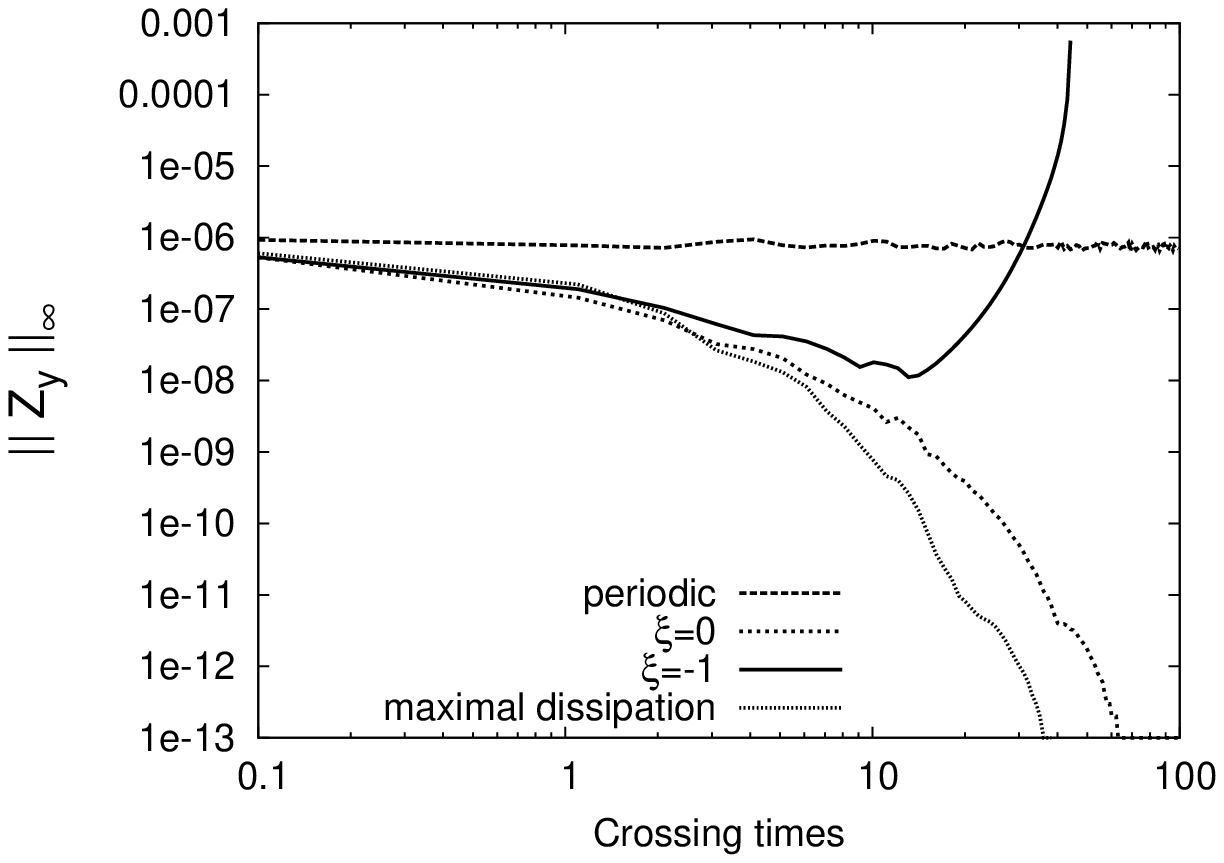}
\caption{Robust stability test. Time evolution of the maximum norm
of the constraint-violating quantities $\Theta$ (left panel), and
$Z_y$ (right panel). The proposed boundary conditions are applied
to all faces. Corner points are avoided in the way shown in
fig.~\ref{stencil}. No extra numerical dissipation has been added,
so that the periodic boundaries plots (dashed lines) keep flat.
The absence of extra dissipation clarifies the features shown in
the previous figure.}\label{robust3D}
\end{figure}

This discretization variant allows getting stable results, at
least for the value $\zeta=0$ of the ordering parameter. We plot
in fig.~\ref{robust3D} the maximum norm of the
constraint-violation quantities $(\Theta,\,Z_i)$. We can see there
that removing the extra numerical dissipation makes the features
more transparent. The instability of the $a_E=2$ choice of the
energy coupling parameter, appears now instantly. The downfall
rate in the stable case $a_E=1$, increased as the constraint
violations are drained out in all three directions now, can be
seen in a more unambiguous way. Concerning the momentum constraint
(right panel), the standard $\zeta=-1$ ordering shows now clearly
its unstable behavior, which was masked by the added dissipation
in the standard treatment (see fig.~\ref{robustOlsson}). The
centered ordering choice $\zeta=0$ recovers instead the manifest
stable behavior shown in single-face simulations (see
fig.~\ref{robust1}), close to the maximal dissipation case (left
panel, bottom line).

Our results show the numerical stability of the proposed boundary
conditions in the linear regime for suitable combinations of the
coupling and/or ordering parameters.  The proposed boundary
conditions produce instead a very effective decreasing of (the
cumulated effect of) energy and momentum constraint violations
which compares with the one obtained by applying maximal
dissipation boundary conditions to (the right-hand-side of) the
constraint related modes.

\section{Gowdy waves as a strong field test}

Although the results of the preceding are encouraging, let us
remark that we were just testing the linear regime around
Minkowsky spacetime. This is not enough, as high-frequency
instabilities can appear in generic, strong field, situations (see
for instance ref.~\cite{ST05}). In order to test the strong field
regime, we will consider the Gowdy solution~\cite{Gowdy71}, which
describes a space-time containing plane polarized gravitational
waves. This is one of the test cases that is used in numerical
code cross-comparison with periodic boundary
conditions~\cite{Mexico,Mexico2}. One of the advantages is that it
allows for periodic and/or reflecting boundary conditions, which
can be applied to the modes which are not in the energy-momentum
constraint sector. A first proposal for this selective testing of
the constraint-related modes has been presented
recently~\cite{bilbao}.

The Gowdy line element can be written as
\begin{equation}\label{gowdy_line}
  {\rm d}s^2 = t^{-1/2}\, e^{Q/2}\,(-{\rm d}t^2 + {\rm d}z^2)
  + t\,(e^P\, {\rm d}x^2 + e^{-P}\, {\rm d}y^2)
\end{equation}
where the quantities $Q$ and $P$ are functions of $t$ and $z$ only
and periodic in $z$, that is~\cite{Mexico},
\begin{eqnarray}
\label{function_P}
  P &=& J_0 (2 \pi t)\; \cos(2 \pi z)
\\
\label{fucntion_L}
  Q &=&  \pi J_0 (2 \pi) J_1 (2 \pi)
   -2 \pi t J_0 (2 \pi t) J_1 (2 \pi t) \cos^2(2 \pi z)
\nonumber \\
  &+& 2 \pi^2 t^2 [{J_0}^2 (2 \pi t)  + {J_1}^2 (2 \pi t)
  - {J_0}^2 (2 \pi)  - {J_1}^2 (2 \pi)]
\end{eqnarray}
so that the lapse function $~\alpha = t^{-1/4}\; e^{Q/4}~$ is
constant in space at any time $t_0$ at which $J_0(2\pi t_0)$
vanishes.

Let us now perform the following time coordinate transformation
\begin{equation}\label{gowdy_time}
  t~=~t_0\;e^{-\tau / \tau_0},
\end{equation}
so that the expanding line element (\ref{gowdy_line}) is seen in
the new time coordinate $~\tau~$ as collapsing towards the $t=0$
singularity, which is approached only in the limit
$\tau\rightarrow\infty$. This 'singularity avoidance' property of
the $~\tau~$ coordinate follows from the fact that the resulting
slicing by $\tau=const$ surfaces is harmonic \cite{BM83}. We will
launch our simulations in normal coordinates, starting with a
constant lapse $\alpha_0=1$ at $\tau=0$ ($t=t_0$).

The discretization is performed like in the preceding section, but
with a space resolution $h=1/100$. Allowing for the fact that the
only non-trivial space dependence in the metric is through $\cos(2
\pi z)$, the numerical grid is fitted to the range $0 \leq z \leq
1$. In this way, the exact solution admits either periodic or
reflection boundary conditions. We can use these exact boundary
conditions as a comparison with the constraint-preserving ones
that we are going to test. As the Gowdy metric components depend
on just one coordinate, we will apply here the proposed
constraint-preserving conditions only to the $z$ faces, keeping
periodic boundary conditions along the transverse directions.
Also, like in the preceding section, we show the results for the
$\,a_E=a_N=a_{\,T} = 1$ coupling parameters combination, although
other choices of $\,a_N,\,a_{\,T} = 1,\,2\,$ lead to similar
results.

\begin{figure}[h]
\centering
\includegraphics[width=8cm, height=6cm]{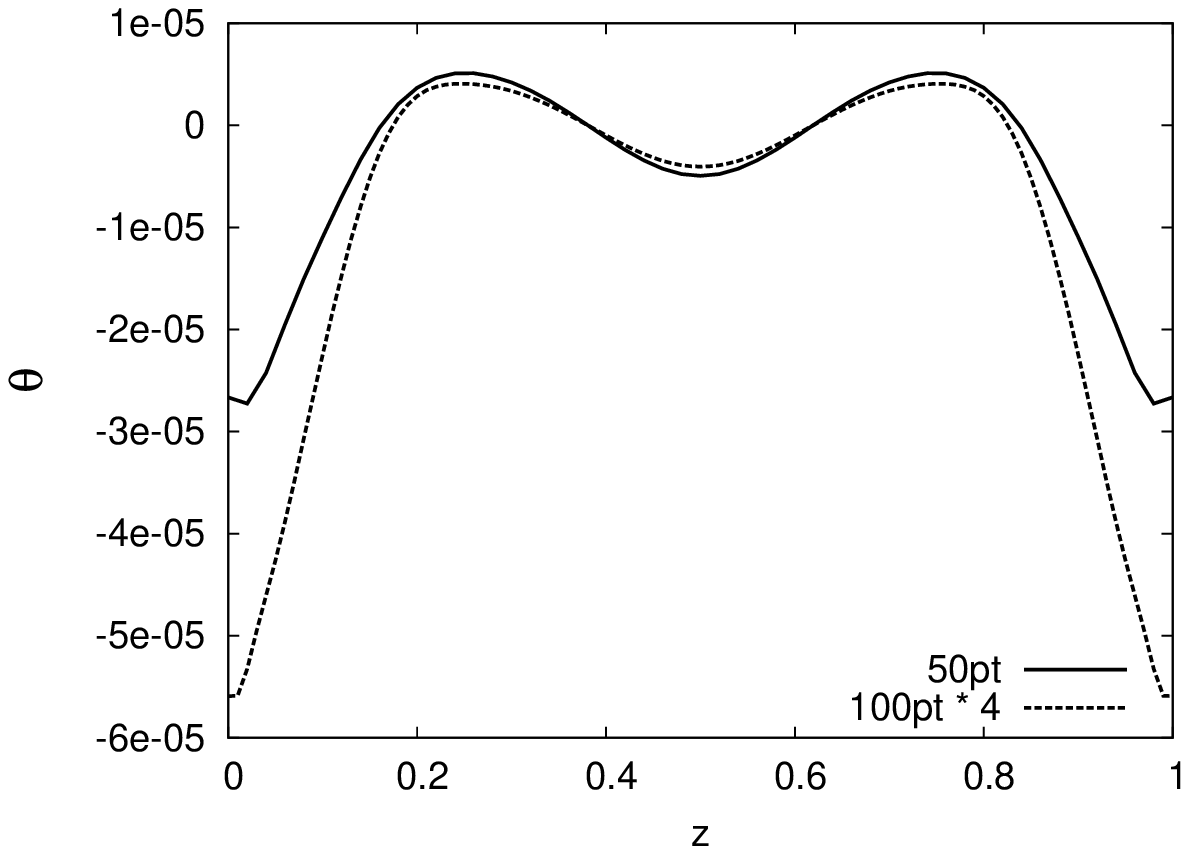}
\includegraphics[width=8cm, height=6cm]{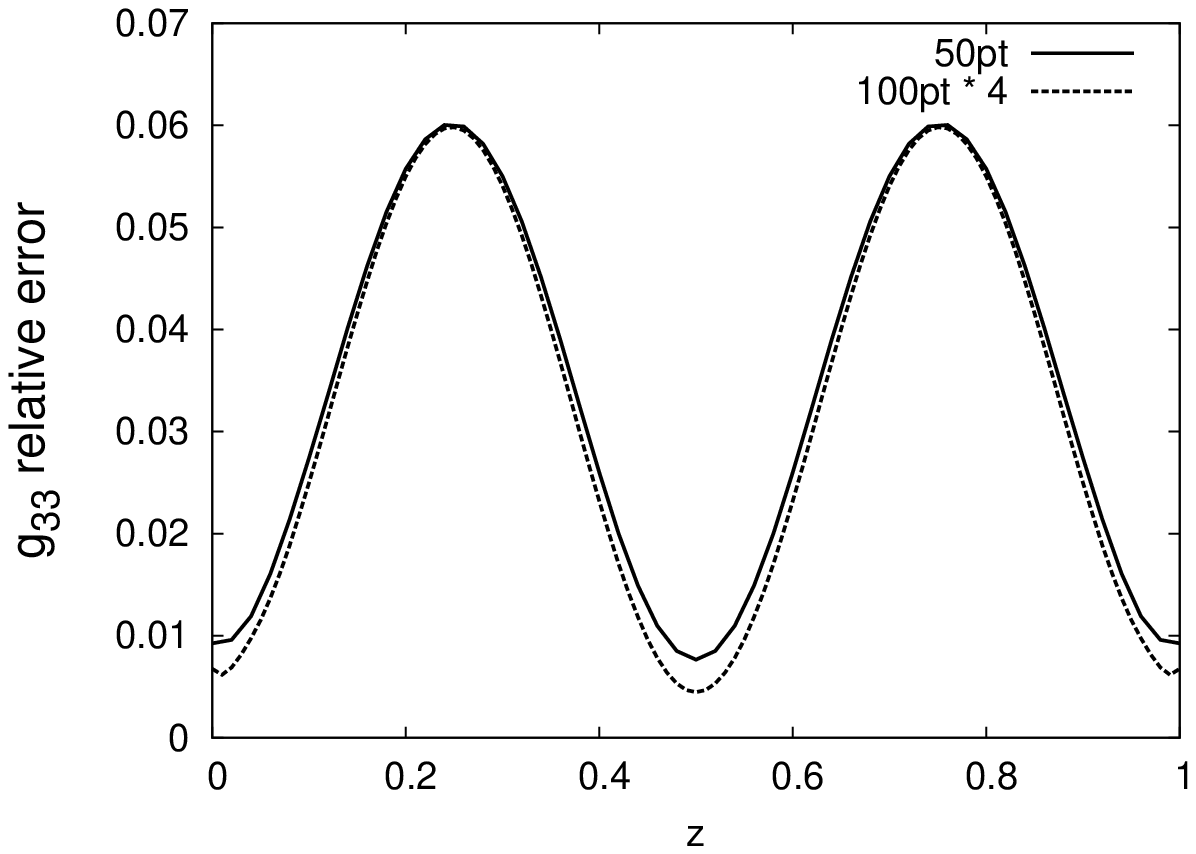}
\caption{Convergence test. The constraint-violating quantity
$\Theta$ is plotted at $t=10$ for two different resolutions (left
panel). We see second-order convergence in the interior region,
decreasing up to first-order at points causally connected to the
boundary, as expected from our SBP algorithm. We plot in the right
panel the relative error of the $g_{zz}$ metric component, evolved
up to $t=250$. The boundary-induced accuracy reduction is not even
visible yet. }\label{Gowdy conv}
\end{figure}

We start with a simple convergence test. As we know the exact
solution (\ref{gowdy_line}), we can directly compute the relative
error of every simulation. Then, only two different resolutions
are required for checking convergence. We will take $h=1/50$ and
$h=1/100$ for our Gowdy wave simulations. We plot in
fig.~\ref{Gowdy conv} the energy-constraint-violation quantity
$\Theta$ at some early time $t=10$ (left panel). We see the
expected second-order accuracy at the interior points which are
yet causally disconnected with the boundaries. We see just
first-order accuracy at the boundary, plus a smooth transition
zone. This accuracy reduction at boundaries is inherent to simple
SBP algorithms, which require a lower-order discretization at
boundary points~\cite{Olsson}. One could keep instead the accuracy
level of the interior points by using more accurate predictions
for boundary values, but at the price of loosing the SBP property.
In our test case, however, this issue is not affecting the metric
components, even at a much later time $t=250$, as we can see in
the right panel of fig.~\ref{Gowdy conv}.

\begin{figure}[h]
\centering
\includegraphics[width=8cm, height=6cm]{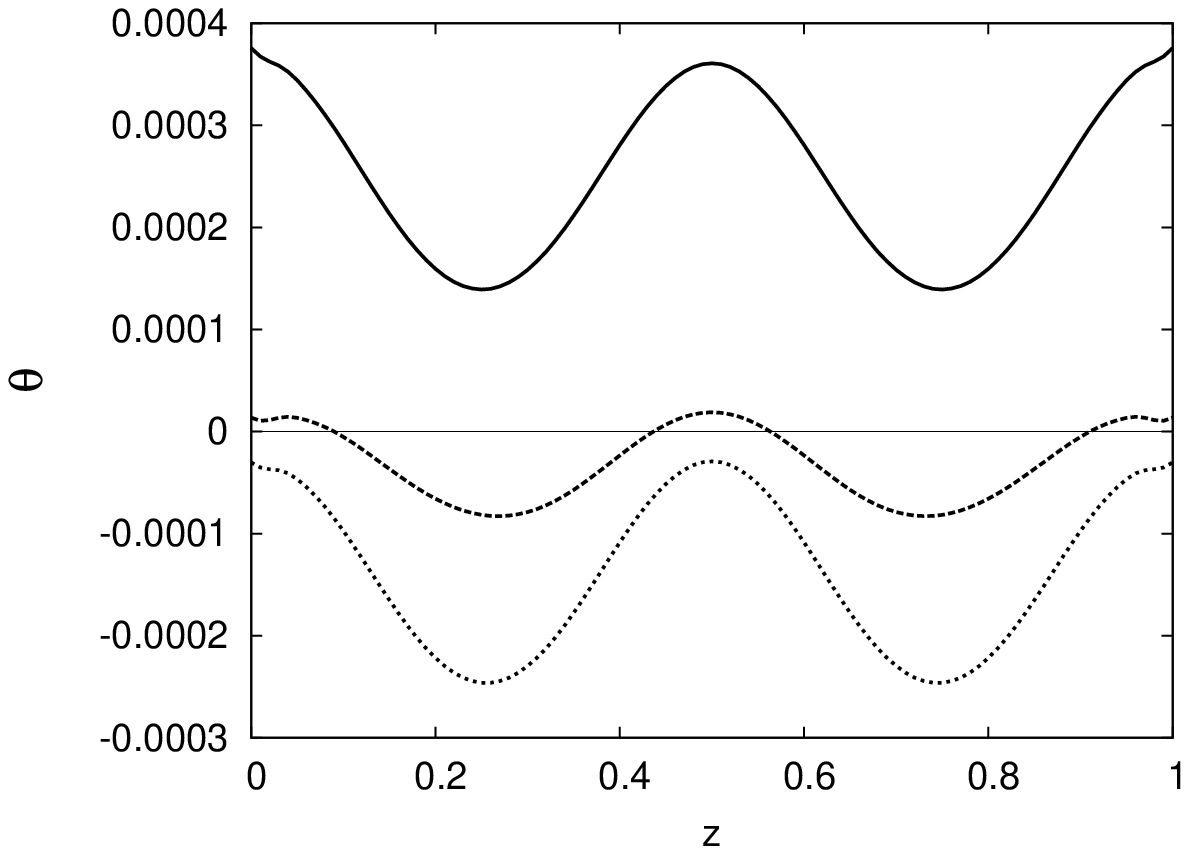}
\includegraphics[width=8cm, height=6cm]{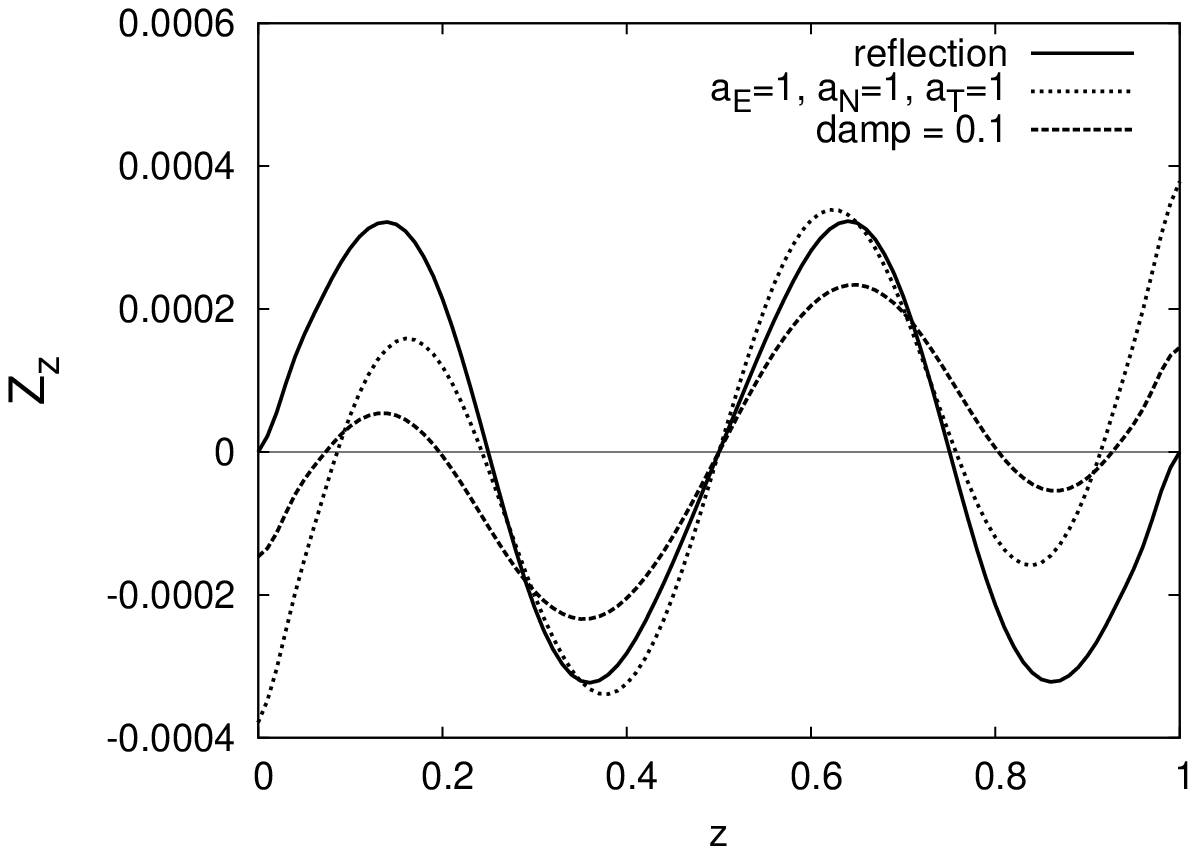}
\caption{Gowdy waves test. The $\Theta$ and $Z_z$ profiles are
plotted as indicators of the accumulated error due to
energy-momentum constraint violations. Reflection boundaries
results are also plot for comparison (continuous lines). Dotted
lines correspond to the proposed boundary conditions, whereas
dashed lines correspond to the same conditions with the extra
damping terms discussed in this section, with
$\eta=0.1$.}\label{GWperiodic}\end{figure}

We show in fig.~\ref{GWperiodic} the results of the $h=1/100$
resolution simulation for the boundary profiles of $\Theta$ and
$Z_i$, indicating the accumulated amount of energy and momentum
constraint violations (left and right panels, respectively). We
apply in both cases the proposed boundary conditions at the $z$
faces to the constraint-related modes, while keeping exact
(reflection) boundary conditions for the other modes. We can see
that the constraint-preserving conditions result, in this
strong-field test, into an accumulated amount of constraint
violations (dotted lines) that is similar or even slightly better
than the one produced by the interior points treatment, which can
be seen in simulations with (exact) reflection boundaries for all
faces (continuous lines). Note that the reflection conditions
anchor $Z_z$ to zero at the boundary points, which is always more
accurate in this test, although not very useful in more realistic
cases.

These results confirm that the proposed boundary conditions are
indeed constraint-preserving, in the sense that their contribution
to energy and momentum constraint violations keeps within the
limits of the truncation error of the discretization algorithms,
even in this strong field scenario. This good behavior can be
further improved by introducing constraint-damping terms in the
evolution equations for the boundary quantities (\ref{Thetabound},
\ref{Zbound}) that is
\begin{eqnarray}\label{damping}
     \dot{\Theta} &+& n^k\Theta_k = -\eta~\Theta\\
     \dot{Z_i} &+& n^kZ_{ki} = -\eta~Z_i\,.
\end{eqnarray}
The resulting values can then be used in the replacements (\ref{E-
repl}) and (\ref{M- repl}), respectively. We have included the
corresponding results in fig.~\ref{GWperiodic} (dashed lines). The
amount of both energy and constraint violations becomes even lower
than the one for the (exact) reflection boundaries simulations
even for a small value $\eta=0.1$ of the damping parameter. The
effect is specially visible in the energy constraint case (left
panel).

\section{Conclusions and outlook}
The work presented in this paper revises and improves the previous
results for the Z4 case~\cite{CP Z4,LNP} in many different ways.

On the theoretical side, we have proposed a new symmetrizer, which
extends the parametric domain for symmetric hyperbolicity from the
single value $\zeta=-1$ to the interval $-1 \leq \zeta \leq 0$. We
have identified in the process a new basis for the dynamical field
space (\ref{Pimu a}-\ref{Pimu d}) which allows a clear-cut
separation between the constraint-related modes and the remaining
ones. Regarding the boundary treatment, we have also generalized
the way in which boundary conditions can by used for modifying the
incoming modes, by introducing a new parameter $a$ which, at least
for the momentum constraint modes, can depart from the standard
value $a=1$ without affecting the stability of the results.

On the numerical side, the use of the new basis definitely
improves the stability of the previous Z4 results. In the single
face case, where we use periodic boundary conditions along
transverse directions, we see that the linear modes previously
reported in the robust stability test~\cite{CP Z4,LNP} for the
symmetric ordering case ($\zeta=0$) no longer show up. Moreover,
we have devised a simple finite-differences stencil for the
prediction step at the boundaries which avoids the corner and
vertex points even in cartesian-like grids, providing an
interesting alternative to the standard (Olsson) corners
treatment.

The proposed boundary conditions have been also tested in a strong
field scenario, the Gowdy waves metric~\cite{Gowdy71}, so that the
effect of non-trivial metric coefficients can be seen in the
simulation results. The convergence test in this non-linear regime
provides strong evidence of numerical stability for some suitable
parameter combinations. Our simulations actually confirm that the
proposed boundary conditions are constraint-preserving: the
accumulated amount of energy-momentum constraint violations is
similar or even better than the one generated by either periodic
or reflection conditions, which are exact in the Gowdy waves case.

Now it remains the question of how these interesting results can
be extended to other 3+1 evolution formalisms and/or gauge
conditions. Let us remember that all our symmetric hyperbolicity
results apply as usual just to the harmonic slicing, not to the
'1+log' class of slicing conditions which are customary in BSSN
black-hole simulations. There is no problem, however, in extending
the proposed boundary conditions to this case: in our new basis
the gauge sector is clearly separated from the constraint-related
one, so that one can keep using the replacements (\ref{E- repl},
\ref{M- repl}) even in this non-harmonic case. The shift, however,
introduces new couplings and would require a detailed case-by-case
investigation: even the strong hyperbolicity of the system can
depend on the specific choice of shift condition.

Concerning the extension from the Z4 to the BSSN formalism, the
momentum constraint treatment can be derived from the simple
condition~\cite{LNP}
\begin{equation}\label{Gamma to Z}
    \widetilde{\Gamma}_i = - \widetilde{\gamma}_{\,ik}\,
    \widetilde{\gamma}^{\,jk}_{~~,j} +2\,Z_i
\end{equation}
which relates the additional BSSN quantity $\widetilde{\Gamma}_i$
with the space vector $Z_i$. The replacement (\ref{M- repl}) can
then be used for getting a suitable boundary condition in this
context. The case of the energy constraint is more challenging, as
the BSSN formalism does not contain any supplementary quantity
analogous to $\Theta$. One could follow, however, the line
recently proposed in~\cite{Z4BSSN}: a slight modification of the
original BSSN equations allows to include the new quantity
$\Theta$, so that the correspondence with the Z4 formalism is
complete. The replacement (\ref{E- repl}) can then be used
directly in such context.

A major challenge is posed by the fact that most BSSN
implementations are of second order in space. This has some
advantages in this context, as the ordering constraints do not
show up and this removes the main source of ambiguities in the
constraint-violations evolution system. As a result, the boundary
conditions (\ref{Thetabound}, \ref{Zbound}) become simply
advection equations so that we can expect a more effective
constraint-violation 'draining' rate. The problem, however, is
that second-order implementations do not have the algebraic
characteristic decomposition which is crucial in the first-order
ones. The boundaries treatment takes quite different approaches in
second-order formalisms, although the evolution equations for the
constraint-related quantities $\Theta\,,~Z_i$ are still of first
order in the Z4 case, even at the continuum level, and this
suggests that the results presented here can be still helpful in
this case. We are currently working in this direction.

\section*{Acknowledgments}
This work has been jointly supported by European Union FEDER funds
and by the Spanish Ministry of Science and Education (projects
FPA2007-60220, CSD2007-00042 and ECI2007-29029-E). C.~Bona-Casas
acknowledges the support of the Spanish Ministry of Science, under
the FPU/2006-02226 fellowship.

\renewcommand{\theequation}{A.\arabic{equation}}
\setcounter{equation}{0}\section*{Appendix A: Positivity of the
energy estimate}

We have derived in section II an generalized '\,energy estimate'
for the Z4 system, namely:
\begin{equation}\label{AE}
    S = \Theta^2 + V_kV^k + \Pi^{ij}\Pi_{ij}
    + \tilde{\mu}^{kij}\tilde{\mu}_{kij}
    + (1+\zeta) (Z^kZ_k-\tilde{\mu}^{kij}\tilde{\mu}_{ijk})
     + 2\, \zeta\, Z_kW^k\,,
\end{equation}
where we noted
\begin{equation}\label{Amut}
    \tilde{\mu}_{kij} = \mu_{kij} - W_k\, \gamma_{ij}\,.
\end{equation}

In order to check the positivity of (\ref{AE}), let us consider
the decomposition of the three-index tensor $~\tilde{\mu}_{kij}~$
into its symmetric and antisymmetric parts, that is
\begin{equation}\label{Amudecomp}
    \tilde{\mu}_{kij} = \tilde{\mu}_{(kij)} +
    \tilde{\mu}^{(a)}_{kij}~.
\end{equation}
Allowing for the identities,
\begin{equation}\label{Amuident}
    \tilde{\mu}^{(a)}_{(kij)}=0\,,\qquad
    \tilde{\mu}^{(a)}_{(ij)k}=-\frac{1}{2}~\tilde{\mu}^{(a)}_{kij}~,
\end{equation}
the rank-three terms contribution to $S$ can the be written as
\begin{equation}\label{A3index}
  S = -\,\zeta~\tilde{\mu}_{(kij)}~\tilde{\mu}^{(kij)} +
  \frac{3+\zeta}{2}~\tilde{\mu}^{(a)}_{kij}~\tilde{\mu}^{(a)}\,^{kij}
  + \cdots
\end{equation}
(the dots stand for lower-rank components). It follows that a
necessary condition for positivity is $~0\,\ge\zeta\,\ge-3$.

We can now rewrite (\ref{AE}) as
\begin{equation}\label{AEbis}
     S = \Theta^2 + V_kV^k + \Pi^{ij}\,\Pi_{ij}
    -\zeta~\tilde{\mu}_{kij}~\tilde{\mu}^{kij}
    + \frac{3}{2}~(1+\zeta)~\tilde{\mu}^{(a)}_{kij}~\tilde{\mu}^{(a)}\,^{kij}
    +  (1+\zeta)~Z^kZ_k + 2\, \zeta\, Z_kW^k.
\end{equation}
Allowing for (\ref{Amut}), which implies in turn
\begin{equation}
    {\tilde{\mu}^k}_{~ki} = - Z_i -W_i\,,
\end{equation}
we see that we can rewrite again (\ref{AEbis}) as
\begin{equation}\label{AEfinal}
     S = \Theta^2 + V_kV^k + \Pi^{ij}\Pi_{ij}
    -\zeta~\tilde{\lambda}_{kij}~\tilde{\lambda}^{kij}
    + \frac{3}{2}~(1+\zeta)~\tilde{\mu}^{(a)}_{kij}~\tilde{\mu}^{(a)}\,^{kij}
    +  (1+\zeta)~Z^kZ_k\,,
\end{equation}
where
\begin{equation}\label{Alambdat}
    \tilde{\lambda}_{kij} = \lambda_{kij}\,|_{\zeta=-1}
    = \tilde{\mu}_{kij} + \gamma_{k(i}W_{j)}\,.
\end{equation}
It follows from the final expression (\ref{AEfinal}) that the
energy estimate is positive definite in the whole interval
\begin{equation}\label{Ainterv}
    0 ~\ge~ \zeta ~\ge~ -1 \,.
\end{equation}
Note that for $\zeta=-1$, that is $\lambda=\tilde{\lambda}$, we
recover the estimate given in ref.~\cite{Z4}.

\renewcommand{\theequation}{B.\arabic{equation}}
\setcounter{equation}{0}
\section*{Appendix B: Strong hyperbolicity of the energy modes}

We can analyze the hyperbolicity of the boundary evolution system,
by considering the characteristic matrix along a generic oblique
direction $\mathbf{r}$, which is related to the normal direction
$\mathbf{n}$ by
\begin{equation}
    \mathbf{r} = \mathbf{n}~cos\varphi + \mathbf{s}~sin\varphi~,
\end{equation}
where we have taken
\begin{equation}
    \mathbf{n}^2 = \mathbf{s}^2 = 1\qquad
    \mathbf{n}\cdot\mathbf{s}=0~.
\end{equation}
The strong hyperbolicity requirement amounts to demand that the
characteristic matrix is fully diagonalizable and has real
eigenvalues (propagation speeds) for any value of the angle
$\varphi$.

In order to compute the characteristic matrix, we will consider
the standard form of (the principal part of) the evolution system
as follows
\begin{equation}\label{fluxcons}
    \partial_t~\mathbf{u} + \alpha\,\partial_r\,\mathbf{F}^r(\mathbf{u})
    = \cdots~,
\end{equation}
where $\mathbf{u}$ stands for the array of dynamical fields and
$\mathbf{F}^r$ is the array of fluxes along the direction
$\mathbf{r}$. We will restrict ourselves here to the Energy-modes
subsystem, which consists in the fields
\begin{equation}\label{uarray}
    \mathbf{u} = (E^+\,,~E^-\,,~V_s\,,~V_{p}~\,)\,
\end{equation}
the index $p$ meaning here a projection along the direction
orthogonal both to $\mathbf{n}$ and $\mathbf{s}$.

It is clear that the two components $V_{p}$ are eigenvectors of
the characteristic matrix with zero propagation speed. The
non-trivial fluxes are then:
\begin{eqnarray}
  F^r(E^+) &=& cos\varphi~E^+ + sin\varphi~ V_{s}\\
  F^r(E^-) &=& (a-1)\,cos\varphi~E^- + (1-a)\,sin\varphi~V_{s}\\
  F^r(V_{s}) &=& \frac{1}{2}~sin\varphi~(E^+ + E^-)\,,
\end{eqnarray}
where we have allowed for the modified evolution equation (\ref{E-
modif}). We can see that the one of the characteristic speeds is
zero and the other two are be given by the solutions of
\begin{equation}\label{speedsE}
    (v-\alpha\,cos\varphi)~(v-(a-1)\,\alpha\,cos\varphi) =
    (1-\frac{a}{2})~\alpha^2~sin^2\varphi~,
\end{equation}
which has real distinct solutions for $~a<2~$. The degenerate case
$~a=2~$ is not diagonalizable. It follows that the boundary
evolution subsystem given by the above fluxes is strongly
hyperbolic for $~a<2~$ and weakly hyperbolic for $~a=2~$.

\newpage
\bibliographystyle{prsty}

\end{document}